\documentclass[prd,aps,a4paper,,twocolumn,eqsecnum,nofootinbib,floatfix]{revtex4}  %

\newif\ifusesec
\usesectrue  
   
\usepackage{graphicx} 
\usepackage{amsmath,amsfonts,amssymb}
\usepackage{color}

\newcommand{\eq}{\eqref}
\newcommand{\be}{\begin{equation}}
\newcommand{\ee}{\end{equation}}
\newcommand{\beq}{\begin{equation}}
\newcommand{\eeq}{\end{equation}}
\newcommand{\bea}{\begin{eqnarray}}
\newcommand{\eea}{\end{eqnarray}}
\newcommand{\g}{\gamma}

\begin{document}

\title{Explicit solution of the gravitational two-body problem \\ at the second post-Minkowskian order}

\author{Donato Bini$^{1,2}$, Thibault Damour$^3$, Andrea Geralico$^1$}
  \affiliation{
$^1$Istituto per le Applicazioni del Calcolo ``M. Picone,'' CNR, I-00185 Rome, Italy\\
$^2$INFN, Sezione di Roma Tre, I-00146 Rome, Italy\\
$^3$Institut des Hautes Etudes Scientifiques, 91440 Bures-sur-Yvette, France
}
\date{\today}

\begin{abstract}
The worldlines (in harmonic coordinates) of two gravitationally interacting massive bodies at the second 
post-Minkowskian order are described in explicit form. Both the conservative case and the radiation-reacted case are considered.
We use our results to check the changes, during scattering, of the individual momenta, as well as of the total
angular momentum. High  post-Newtonian order values of the $O(G^2)$ radiation-reaction acceleration components are provided for checks
of future post-Newtonian computations. 
 \end{abstract}

\maketitle
\section{Introduction}

We consider here the gravitational interaction of two (non spinning) massive bodies within a post-Minkowskian (PM) context,
i.e., as an expansion in powers of Newton's constant $G$, without making any small-velocity assumption. Explicit expressions
for the accelerations $\frac{d u_a(\tau_a)}{d \tau_a}=\frac{d^2 z_a(\tau_a)}{d \tau_a^2}$, of the
two worldlines (labelled by $a= A,B$), have been obtained long ago, in harmonic coordinates, at the second post-Minkowskian (2PM) order
\cite{Westpfahl:1969ea,Westpfahl:1979gu,Westpfahl:1980mk,Bel:1981be}.  These 2PM-accurate (i.e. $O(G^2)$ equations of motion have been used 
 to derive the corresponding $O(G^2)$ integrated changes, during scattering, of the individual momenta \cite{Westpfahl:1985tsl}, 
 as well as that of the total angular momentum \cite{Bini:2022wrq}. However, no explicit 2PM-accurate solutions for the two worldlines,
 $z_a^\alpha=z_a^\alpha(\tau_a)$, in terms of their corresponding (Minkowskian) proper times, $\tau_a$, have been
 yet derived. [See, however, Ref. \cite{Bini:2018ywr} for  explicit 1PM-accurate worldlines.]
  Recent years have witnessed spectacular
 progress in the PM knowledge of the integrated changes, during scattering, of the individual momenta of a two-body
 system (e.g. \cite{Bern:2019crd,Bern:2021dqo,Herrmann:2021tct,Bjerrum-Bohr:2021din,Bini:2021gat,Bern:2021yeh,Dlapa:2021vgp,Dlapa:2022lmu,Bini:2022enm,Damgaard:2023ttc}), but no results have been published concerning the equations of
 motion (nor their solution worldlines) beyond the 2PM level.  The aim of the present paper is to provide  explicit 
 worldline solutions at the 2PM accuracy.

In the PM context, the worldlines  ${\mathcal L}_a, (a=A,B)$ (with parametric equations $z_a^\alpha=z_a^\alpha(\tau_a)$) 
are thought of as belonging to a   Poincar\'e-Minkowski spacetime  with metric
\beq
ds^2= \eta_{\alpha\beta}dx^\alpha dx^\beta\,,\qquad \eta_{\alpha\beta}={\rm diag}[-1,1,1,1]\,.
\eeq
The parameter $\tau_a$ is the Minkowski proper time along  ${\mathcal L}_a$: i.e. $\eta_{\alpha\beta}u_a^\alpha u_a^\beta=-1$, where $ u_a(\tau_a)=\frac{d z_a(\tau_a)}{d \tau_a}$. [We use units where $c=1$.]
We work in harmonic coordinates and shall use as starting point the explicit 2PM equations of motion of Ref. \cite{Bel:1981be}.
We generally follow the notation and conventions of the latter reference, except that we label the two bodies by $A$ 
(with mass $m_A$) and $B$ (with mass $m_B$).

The acceleration of body A, due to the gravitational interaction with body B, reads
\beq
\label{acc_terms}
\frac{du_A^\alpha}{d\tau_A} = G\Gamma_{A 1\epsilon}^\alpha(\tau_A, \ldots)+G^2 \Gamma_{A 2\epsilon}^\alpha(\tau_A, \ldots)+O(G^3)\,,
\eeq
where the second subscript, $1$ or $2$, denotes the PM order, while the additional subscript, $\epsilon$, denotes
the retarded ($\epsilon= +1$) or advanced ($\epsilon=-1$) character of the interaction. [The equations of motion
of \cite{Bel:1981be} were only written for the retarded case. The advanced case is obtained by an overall time reversal.]
In our integration below of the $A$ equations of motion, we shall, for simplicity, drop the first, $A$, subscript on the accelerations
 $\Gamma_{A 1\epsilon}^\alpha$,  $\Gamma_{A 2\epsilon}^\alpha$.

As indicated in Eq. \eq{acc_terms}  each contribution on the right-hand side (rhs) is initially given not
as an explicit function of $\tau_A$, but rather as an expression depending on  $z_A(\tau_A)$ and 
also on various other auxiliary quantities
 defined from  $z_A(\tau_A)$ and from the geometry of the two worldlines (such as the  point on ${\mathcal L}_B$ which is retarded, or advanced, 
with respect to $z_A(\tau_A)$; see below). In the following, we focus on explicitly solving Eq. \eq{acc_terms} in terms
of initial data in the asymptotic past, and notably the two incoming four-velocities $u_{A-}$ and $u_{B-}$.
Because of the formal symmetry obtained by swapping the labels $A$ and $B$, it is enough to solve the $A$ equation
of motion Eq. \eq{acc_terms}.

We shall solve Eq. \eq{acc_terms}  perturbatively (in $G$), i.e.
\beq
u_A^\alpha(\tau)=u_{A-}^\alpha+G u_{A1}^\alpha(\tau_A) +G^2 u_{A2}^\alpha(\tau_A)+ O(G^3)\,,
\eeq
with the boundary conditions, $\lim_{\tau_A \to -\infty} u_{A1}^\alpha(\tau_A)= \lim_{\tau_A \to -\infty} u_{A2}^\alpha(\tau_A)= 0$, and
\beq
z_A^\alpha(\tau)=b_A^\alpha +u_{A-}^\alpha \tau_A +G z_{A1}^\alpha(\tau_A, s) +G^2 z_{A2}^\alpha(\tau_A)+ O(G^3)\,,
\eeq
with the  boundary condition, $ \lim_{\tau_A \to -\infty} z_{A2}^\alpha(\tau_A) = 0$. There is a subtlety in
the boundary condition for the 1PM term $z_{A1}^\alpha(\tau_A)$ because it contains a logarithmic drift,
generally parametrized by an arbitrary length scale $s$. We shall show below how to choose the two
 vectorial impact parameters, $b_A^\alpha$, $b_B^\alpha$, as well as the scale $b$ (taken to be $b \equiv |b_A^\alpha -b_B^\alpha|$) so as to get a uniquely fixed solution. Radiation-reaction effects enter already at order $O(G^2)$. We shall
 separately consider below  conservative, and radiation-reacted motions.

\section{Setup} \label{setup}

As we are considering nonspinning bodies, their motion is planar, i.e. the two worldlines are contained in a 3-plane spanned
by three 4-vectors: the two incoming 4-velocities, $u_{A-}$, and $u_{B-}$, and a vectorial impact parameter $b_{AB} \equiv b_A- b_B$. By definition, the latter is orthogonal to both incoming velocities: $u_{A-} \cdot b_{AB}= u_{B-} \cdot b_{AB}=0$.
The construction below of the worldline solutions will ensure that the norm  $b \equiv |b_{AB}|$ is equal to the usually
defined incoming  impact parameter.

When studying the worldline of body $A$, it is convenient to use an {\it orthogonal} basis in the motion 3-plane made of the
following three  4-vectors 
\be \label{basis}
u_{A-}\; ; \; \hat b_{AB} \equiv \frac{b_{AB}}{b} \; ; \; u_{B\perp A}\equiv P(u_{A-})u_{B-}= u_{B-}- \g u_{A-} .
\ee
Here,   $v_{\perp u}  \equiv P(u) v$ denotes the projection of $v$ orthogonally to the unit timelike vector $u$, involving
the projection operator $P(u)^\mu_\nu \equiv \delta^\mu_\nu + u^\mu u_\nu$. We also denoted the incoming relative
Lorentz factor as $\g \equiv - u_{A-} \cdot  u_{B-}$.
The squared norms of the three mutually orthogonal vectors, Eq. \eq{basis}, are: $u_{A-}^2=-1$, $ \hat b_{AB}^2=+1$ and
$ u_{B\perp A}^2= \g^2-1$.

We denote the components of  the four velocity  $u_A^\alpha (\tau_A)$ on the asymptotic basis  \eq{basis}
as,
\bea
\label{u_A_con_Cs}
u_A^\alpha (\tau_A)  &=& C_{u_A}(\tau_A) u_{A-}^\alpha+C_{\hat b_{AB}}(\tau_A) \hat b_{AB}^\alpha\nonumber\\
&+& C_{B\perp A}(\tau_A)u_{B\perp A}^\alpha\,.
\eea
Here, as we shall focus on the $A$ worldline, we dropped for simplicity an extra label $A$ on the coefficients on the rhs.
Each coefficient is expanded in powers of $G$ as
\bea
C_{u_A}(\tau_A)&=& 1+G C_{u_A}^{G^1}(\tau_A)+ G^2 C_{u_A}^{G^2}(\tau_A)+ O(G^3)\,,\nonumber\\
C_{\hat b_{AB}}(\tau_A)&=& G C_{\hat b_{AB}}^{G^1}(\tau_A)+G^2 C_{\hat b_{AB}}^{G^2}(\tau_A)+ O(G^3)\,, \nonumber\\
C_{B\perp A}(\tau_A)&=& G C_{B\perp A}^{G^1}(\tau_A)+G^2 C_{B\perp A}^{G^2}(\tau_A)+ O(G^3)\,.\nonumber\\
\eea
The $G$-expansion of the normalization condition $u_A \cdot u_A=-1$ yields
\bea
\label{eq_norm_u}
C_{u_A}^{G^1}(\tau_A)&=&0\,, \nonumber\\
C_{u_A}^{G^2}(\tau_A) &=&\frac12 \left[ \left( C_{\hat b_{AB}}^{G^1}(\tau_A)\right)^2 +\left(C_{B\perp A}^{G^1}(\tau_A)\right)^2 \right]\,,\qquad
\eea
implying that $C_{u_A}^{G^2}(\tau_A)$ is fully determined by solving the first order motion.

The $C$ coefficients in Eq. \eq{u_A_con_Cs} have also an expansion in powers of the two masses $m_A$ and $m_B$, namely
(remembering that we are considering the worldline of body $A$ as deflected by the gravitational effect of body $B$)
\bea
\label{coeffsG1}
C_{\hat b_{AB}}^{G^1}(\tau_A)&=&  m_B   C_{\hat b_{AB}}^{G^1 m_B}(\tau_A)\,, \nonumber\\
C_{B\perp A}^{G^1}(\tau_A)&=&  m_B   C_{B\perp A}^{G^1 m_B}(\tau_A)\,,
\eea
at the first order, and
\bea
\label{coeffsG2}
C_{u_A}^{G^2}(\tau_A)&=&m_B^2C_{u_A}^{G^2m_B^2}(\tau_A)
\,,\nonumber\\
C_{\hat b_{AB}}^{G^2}(\tau_A)&=& m_B m_A  C_{\hat b_{AB}}^{G^2m_Am_B}(\tau_A) + m_B^2 C_{\hat b_{AB}}^{G^2m_B^2}(\tau_A)
\,,\nonumber\\
C_{B\perp A}^{G^2}(\tau_A)&=&  m_B m_A  C_{B\perp A}^{G^2 m_A m_B}(\tau_A)+ m_B^2  C_{B\perp A}^{G^2m_B^2}(\tau_A)
\,,\nonumber\\
\eea
at the second order.
 
Summarizing, the knowledge of the 2PM-accurate expansion of  $u_A(\tau_A)$ is embodied in six functions of $\tau_A$:
two at $O(G)$,  $C_{\hat b_{AB}}^{G^1m_B}(\tau_A)$ and $ C_{B\perp A}^{G^1m_B}(\tau_A)$; and four
at $O(G^2)$: $ C_{\hat b_{AB}}^{G^2m_Am_B}(\tau_A)$,  $ C_{\hat b_{AB}}^{G^2m_B^2}(\tau_A)$,
$C_{B\perp A}^{G^2 m_A m_B}(\tau_A)$ and $ C_{B\perp A}^{G^2m_B^2}(\tau_A)$.
All these functions must satisfy the boundary condition that they vanish as $\tau_A \to -\infty$, so that 
\beq
\lim_{\tau_A\to -\infty} u_A^\alpha (\tau_A)= u_{A-}^\alpha\,.
\eeq

Finally, after having obtained the parametric solution for  $ u_A (\tau_A)$, one obtains the parametric equation
$ z_A^\alpha (\tau_A)$ of the world line of body $A$ by integrating
\bea
\frac{ d z_A^\alpha (\tau_A)}{d\tau_A}&=&  u_A^\alpha (\tau_A) \nonumber\\
&=& C_{u_A}(\tau_A) u_{A-}^\alpha+C_{\hat b_{AB}}(\tau_A) \hat b_{AB}^\alpha \nonumber\\
&+& C_{B\perp A}(\tau_A)u_{B\perp A}^\alpha\,,
\eea
with some past boundary conditions.

Though all our results for the various coefficients 
$C_{u_A}(\tau_A), C_{\hat b_{AB}}(\tau_A), C_{B\perp A}(\tau_A)$ will be explicitly Lorentz covariant, we will sometimes
find convenient to introduce the following shorthand notation for a frame $(e_t^A,e_x,e_y^A,e_z)$ associated with the
worldline of body $A$:
\beq
e_t^A \equiv u_A^-\,, 
\eeq
\be
e_x \equiv {\hat {\mathbf  b}}_{AB}\equiv  \frac{ {\mathbf b}_{AB}}{||b_{AB}||}=\frac{{\mathbf b}_{AB}}{b},
\ee
\beq
e_y^A \equiv -  \frac{u_{B\perp A}}{\sqrt{\gamma^2-1}}\,,
\eeq
with $e_z$ being orthogonal to the plane of the motion.
In this notation, we have 
\beq
u_B^-=\gamma e_t^A -\sqrt{\gamma^2-1}e_y^A\,.
\eeq
Note that only $e_t^A$ and $e_y^A$ depend on the choice of body $A$ rather than body $B$.

\section{Solution at order $G^1$}

The solutions for  $C_{\hat b_{AB}}^{G^1m_B}(\tau_A)$ and $ C_{B\perp A}^{G^1m_B}(\tau_A)$ 
are uniquely fixed by the boundary conditions at $\tau_A \to -\infty$ and are given (in agreement with Ref. \cite{Bini:2018ywr})
by
\bea
 C_{\hat b_{AB}}^{G^1m_B}(\tau_A)&=& \frac{(1-2\gamma^2)S(\tau_A)}{\sqrt{\gamma^2-1}D(\tau_A)}
\,,\nonumber\\
 C_{B\perp A}^{G^1m_B}(\tau_A)&=& \frac{\gamma (2\gamma^2-3)}{(\gamma^2-1)D(\tau_A)}  \,,
\eea
where we defined
\bea
\label{D_and_S_defs}
D(\tau_A)&\equiv& \sqrt{b^2 +\tau_A^2 (\gamma^2-1)}\,,\nonumber\\
S(\tau_A)&\equiv& \frac{1}{b}\left(\tau_A \sqrt{\gamma^2-1}+D(\tau_A)  \right)\,.
\eea
Let us notice that  when $\tau\to -\infty$, one has $D(\tau_A) \simeq |\tau_A|\sqrt{\gamma^2-1}$, $S(\tau_A)\to 0$, and
\bea
\lim_{\tau_A\to -\infty} C_{\hat b_{AB}}^{G^1 m_B}(\tau_A)&=& O\left(\frac{1}{\tau_A^2}\right)\,, \nonumber\\
\lim_{\tau_A\to -\infty} C_{B\perp A}^{G^1m_B}(\tau_A)&=&  O\left(\frac{1}{|\tau_A|}\right)  \,.
\eea
The $ O\left(\frac{1}{|\tau_A|}\right)$ behavior of the ``longitudinal" component $C_{B\perp A}^{G^1m_B}(\tau_A)$
of $ u_A (\tau_A)- u_{A-}$ implies that the corresponding component of  $ z_A (\tau_A)- u_{A-} \tau_A - b_A$ 
contains a logarithmic drift.

When $\tau_A=0$, $D(\tau_A)$ takes its minimum value $D(0)= D_{\rm min}=b$, while $S(0)=1$, so that
\bea
C_{\hat b_{AB}}^{G^1m_B}(0)&=& \frac{(1-2\gamma^2) }{\sqrt{\gamma^2-1}}\, \frac{1}{b}\,,\nonumber\\
C_{B\perp A}^{G^1m_B}(0)&=& \frac{\gamma (2\gamma^2-3)}{(\gamma^2-1)}\, \frac{1}{b}\,.
\eea 
Finally, when $\tau\to +\infty$, one has $D(\tau_A) \sim \tau_A\sqrt{\gamma^2-1}$ and $S(\tau_A)\to 2 \tau_A\sqrt{\gamma^2-1}/b $, so that
\bea
\lim_{\tau_A\to +\infty} C_{\hat b_{AB}}^{G^1 m_B}(\tau_A)&=& \frac{2(1-2\gamma^2)}{\sqrt{\gamma^2-1}}\frac{1}{b}\,, \nonumber\\
\lim_{\tau_A\to +\infty} C_{B\perp A}^{G^1m_B}(\tau_A)&=&  O\left(\frac{1}{\tau_A}\right)  \,.
\eea
The nonzero value of $\lim_{\tau_A\to +\infty} C_{\hat b_{AB}}^{G^1 m_B}(\tau_A)$ embodies the 1PM-accurate
scattering angle.

Let us now pass to the parametric equations of the world lines which are obtained by integrating
\bea
\frac{ d z_A^\alpha (\tau_A)}{d\tau_A} &=& C_{u_A}(\tau_A) u_{A-}^\alpha+C_{\hat b_{AB}}(\tau_A) \hat b_{AB}^\alpha \nonumber\\
&+& C_{B\perp A}(\tau_A)u_{B\perp A}^\alpha\,.
\eea

We have seen  that $ C_{u_A}(\tau_A)= 1+ O(G^2)$, and that, in the asymptotic past, $C_{\hat b_{AB}}(\tau_A) = O\left(\frac{1}{|\tau_A|^2}\right)$ so that one can uniquely integrate the differential equation
for the components of  $ z_A^\alpha (\tau_A)$ along $u_{A-}^\alpha$ and along the transverse direction $\hat b_{AB}$,
with the past boundary conditions discussed above.
By contrast the component of  $ z_A^\alpha (\tau_A)$ along the direction $u_{B\perp A}$ involves the logarithmically divergent integral 
\beq
\int^{\tau_A}  d\tau_A'   C_{B\perp A}(\tau_A')\,.
\eeq
We  therefore need to introduce a scale, say $s$, and work with the indefinite integral
\bea \label{indef_int}
\int^{\tau_A}  d\tau_A'    C_{B\perp A}^{G^1m_B}(\tau_A') &=&\frac{\gamma (2\gamma^2-3)}{(\gamma^2-1)^{3/2}}\times \nonumber\\
&& \ln \left(\frac{\tau_A \sqrt{\gamma^2-1}+D(\tau_A)}{s} \right)
\,.\nonumber\\
\eea 
We could continue the integration at higher PM levels by keeping arbitrary the scale $s$. An infinitesimal change 
in the choice of $s$ can be absorbed in correlated shifts in $b_A$, $b_B$ and  in the proper times $\tau_A$ and $\tau_B$,
and has therefore no intrinsic physical content. Here, we choose to fix the scale  as $s=b$, 
the natural length scale entering the problem. This replaces the indefinite integral \eq{indef_int} by the following
definite one
\bea
\label{def_int}
&&\frac{\gamma (2\gamma^2-3)}{(\gamma^2-1)^{3/2}} \ln \left(\frac{\tau_A \sqrt{\gamma^2-1}+D(\tau_A)}{b} \right)  \nonumber\\
&&=\frac{\gamma (2\gamma^2-3)}{(\gamma^2-1)^{3/2}} \ln \left(S(\tau_A)\right) \nonumber\\
&& = \int^{\tau_A}_0  d\tau_A'   C_{B\perp A}^{G^1 m_B}(\tau_A')\,.
\eea
This finally yields the following specific $O(G)$-accurate solution
\bea
z_A^\alpha(\tau_A)&=& b_A^\alpha +u_A^-\tau_A+ G m_B {\mathcal I}^{G^1 m_B}_{\hat b_{AB}}(\tau_A)\hat b_{AB}^\alpha\nonumber\\
&+& G m_B  {\mathcal I}^{G^1 m_B}_{B\perp A}(\tau_A)u_{B\perp A}^\alpha  +O(G^2)\,,
\eea
where
\bea
\label{intCG1}
{\mathcal I}^{G^1 m_B}_{\hat b_{AB}}(\tau_A)&=&\int_{-\infty}^{\tau_A}d\tau_A'   C_{\hat b_{AB}}^{G^1m_B}(\tau_A')\nonumber\\
&=&\frac{1-2\gamma^2}{\gamma^2-1}S(\tau_A)\,, \nonumber\\
{\mathcal I}^{G^1 m_B}_{B\perp A}(\tau_A)&=&\int_{0}^{\tau_A}d\tau_A'  C_{B\perp A}^{G^1 m_B}(\tau_A')\nonumber\\
&=& \frac{\gamma (2\gamma^2-3)}{(\gamma^2-1)^{3/2}} \ln S(\tau_A)
\,.\nonumber\\
\eea
The position of the first worldline at the time $\tau_A=0$ is
\beq
z_A^\alpha(0)=b_A^\alpha + G m_B \frac{1-2\gamma^2}{\gamma^2-1}\hat b_{AB}^\alpha  +O(G^2)\,.
\eeq
Exchanging $A$ and $B$ (and therefore swapping $b_{AB}^\alpha \to b_{BA}^\alpha = -b_{AB}^\alpha$) yields
\beq
z_B^\alpha(0)=b_B^\alpha - G m_A \frac{1-2\gamma^2}{\gamma^2-1}\hat b_{AB}^\alpha\,,
\eeq
so that
\bea
z_A^\alpha(0)-z_B^\alpha(0)&=& b_A^\alpha-b_B^\alpha + G (m_A+m_B) \frac{1-2\gamma^2}{\gamma^2-1}\hat b_{AB}^\alpha\nonumber\\
&=& \left(b+G (m_A+m_B) \frac{1-2\gamma^2}{\gamma^2-1}\right) \hat b_{AB}^\alpha\,. 
\nonumber\\
\eea
The present 1PM-accurate solution differs from the one written down in \cite{Bini:2018ywr} by  shifts in $b_A^\alpha $
and $b_B^\alpha $, namely 
$b_A^{\rm there} = z_A^{\rm here}(0)=b_A^{\rm here} +  G m_B \frac{1-2\gamma^2}{\gamma^2-1}\hat b_{AB}$
and $b_B^{\rm there} = z_B^{\rm here}(0)= b_B^{\rm here} -  G m_A \frac{1-2\gamma^2}{\gamma^2-1}\hat b_{AB}$. As a consequence
the ``bare" impact parameter $b_0$ used in \cite{Bini:2018ywr} is equal to
$b_0 = b+G (m_A+m_B) \frac{1-2\gamma^2}{\gamma^2-1}+O(G^2)$.

\section{Advanced and retarded quantities}

The 2PM equations of motion derived in Ref. \cite{Bel:1981be} were written in terms of intermediate geometric quantities
defined on the two worldlines by means of several connecting null rays. In particular, the null cone emanating from
 from some point $z_A(\tau_A)$ on ${\mathcal L}_A$ will intersect  ${\mathcal L}_B$ 
in two points corresponding to the two values of $\tau_B$ solving  the equation
\beq
\eta_{\alpha\beta}(z_A^\alpha(\tau_A)-z_B^\alpha(\tau_B))(z_A^\beta(\tau_A)-z_B^\beta(\tau_B))=0\,.
\eeq  
The retarded solution will be denoted  $\tau_{B,\rm ret}(\tau_A)$, and the advanced one $\tau_{B,\rm adv}(\tau_A)$. 
These solutions are obtained by working perturbatively in $G$. As already said, we shall use the indicator $\epsilon$ 
to distinguish these two solutions, with $\epsilon=+1$ corresponding to the retarded case, 
and $\epsilon=-1$ corresponding to the advanced one. 
For example, $u_B(\tau_{B,\rm ret/adv}(\tau_A))$ and  $z_B(\tau_{B,\rm ret/adv}(\tau_A))$ will be denoted as $u_{B,\epsilon}(\tau_A)$ and $z_{B,\epsilon}(\tau_A)$, respectively.

The following retarded/advanced vectorial and scalar quantities will be needed to express the acceleration terms entering Eq. \eqref{acc_terms} (we recall that  $P(u)=\eta + u\otimes u$ projects orthogonally to $u$)
\bea
\omega_\epsilon &=&u_A  \cdot  u_{B,\epsilon}\,,\nonumber\\ 
\rho_\epsilon &=& -\epsilon (z_A- z_{B,\epsilon})\cdot  u_{B,\epsilon}\,, \nonumber\\
{\boldsymbol \nu}_\epsilon &=&  
\frac{1}{\rho_\epsilon}P(  u_{B,\epsilon})(z_A- z_{B,\epsilon})\,,\nonumber\\
{\mathbf A}_\epsilon &=&\frac{1}{\rho_\epsilon}P(u_A)[z_A-  z_{B,\epsilon}]\,, \nonumber\\
 A_\epsilon &=&\sqrt{{\mathbf A}_\epsilon\cdot {\mathbf A}_\epsilon}\,,\nonumber\\
{\mathbf v}_\epsilon&=&P(u_A)u_{B,\epsilon}\,.
\end{eqnarray}
The  decomposition of the null vector  $z_A-  z_{B,\epsilon}$ with respect to $u_A$ can be written in two alternative ways
\bea
z_A-  z_{B,\epsilon}= \rho_\epsilon ({\mathbf A}_\epsilon +u_A  \epsilon  A_\epsilon )=B_\epsilon ({\mathbf N}_\epsilon +u_A  \epsilon   )
\eea
where 
\be
 B_\epsilon=\rho_\epsilon A_\epsilon\,,
 \ee
 and where
 \be
 {\mathbf N}_\epsilon = \frac{{\mathbf A}_\epsilon}{A_\epsilon }
 \ee
 is the unit spatial vector in the direction of ${\mathbf A}_\epsilon$. The various explicit powers of $\epsilon$ entering
 the above equations are related to the fact that the component of $(z_A-  z_{B,\epsilon})$ along $u_A$ is positive
 for the retarded case, and negative for the advanced case.

The vectorial and scalar quantities needed to express the acceleration at the lowest order are listed below (using the
notation introduced at the end of Sec. \ref{setup})
\bea
\tau_{B,\epsilon}&=&\gamma \tau_A - \epsilon D(\tau_A) + O(G)
\,,\nonumber\\
\omega_\epsilon&=&-\gamma + O(G)
\,,\nonumber\\
\rho_\epsilon&=&D(\tau_A) + O(G)
\,,\nonumber\\
(z_A-  z_{B,\epsilon})(\tau_A)&=& be_x +\tau_A u_A-\tau_{B,\epsilon} u_B + O(G)
\,,\nonumber\\
{\mathbf A}_\epsilon&=&\frac{1}{D(\tau_A)}[b e_x +(\gamma \tau_A -\epsilon D(\tau_A)) \sqrt{\gamma^2-1}e_y^A]\nonumber\\ 
&+& O(G)
\,,\nonumber\\
A_\epsilon&=&\gamma -\epsilon \frac{\gamma^2-1}{D(\tau_A)}\tau_A + O(G)
\,.
\eea
Note also the relations
$\frac{D(\tau_{B,\epsilon} )}{D(\tau_A)}=A_\epsilon(\tau_A)$, $\frac{S(\tau_{B,\epsilon})}{S(\tau_A)}=\gamma-\epsilon \sqrt{\gamma^2-1}$, so that
$\ln S(\tau _{B,\epsilon})=\ln S(\tau_A)+\ln (\gamma-\epsilon \sqrt{\gamma^2-1})=\ln S(\tau_A)-\ln (\gamma+\epsilon \sqrt{\gamma^2-1})=\ln S(\tau_A)-\epsilon\, {\rm arccosh}(\gamma)$.

\section{Solution at order $G^2$}

The 2PM-accurate equations of motion of Ref. \cite{Bel:1981be} are not given as explicit $G$-expanded functions of $\tau_A$, of the form
\bea
\label{accexpl0}
\frac{d u_A^\alpha(\tau_A)}{d\tau_A}&=&G A_A^{1\alpha}(\tau_A)+G^2 A_{A\,\epsilon}^{2\alpha}(\tau_A) +O(G^3)\,,\qquad
\eea
but they are   expressed in terms of a set of (retarded or advanced) intermediate quantities such as
\be
X_\epsilon=[\rho_\epsilon, \omega_\epsilon, A_\epsilon, {\mathbf A}_\epsilon, {\mathbf v}_\epsilon]
\ee
 Therefore, when starting from the 2PM-accurate equations of motion of body $A$ from Ref. \cite{Bel:1981be}, written as
 (where we recall that we drop the extra label $A$ on the rhs of these equations)
 \bea
\label{acc_sec_ord}
\frac{du_A^\alpha(\tau_A)}{d\tau_A} &=& G\Gamma_{1\epsilon}^\alpha[X_\epsilon] \nonumber\\
&+& G^2 \Gamma_{2\epsilon}^\alpha[X_\epsilon] +O(G^3)\,,
\eea
we must insert in them the $G$-expansion of the various arguments $X_\epsilon$ considered as explicit
functions of $\tau_A$, say
\beq
X_\epsilon(z_A, z_{B \epsilon}, \ldots)=X_\epsilon^0(\tau_A)+G X_\epsilon^1(\tau_A)+\ldots\, .
\eeq
At the $G^2$ accuracy, it is enough to use the 1PM-accurate value of $X_\epsilon$ in $G \Gamma_{1\epsilon}[X_\epsilon]$,
and the  leading-order (straight line) value of $X_\epsilon$ in $G^2 \Gamma_{2\epsilon}[X_\epsilon]$, say
\bea
\label{acc_sec_ord_G}
\frac{du_A^\alpha(\tau_A)}{d\tau_A} &=& G\Gamma_{1\epsilon}^\alpha[X_\epsilon^0+G X_\epsilon^1](\tau_A)\nonumber\\
&+& G^2 \Gamma_{2\epsilon}^\alpha[X_\epsilon^0](\tau_A) +O(G^3)\,,
\eea

The first-order acceleration, expressed in terms of
$X_\epsilon=[\rho_\epsilon, \omega_\epsilon,A_\epsilon, {\mathbf A}_\epsilon, {\mathbf v}_\epsilon]$, reads
\bea
\label{Gamma1_functional}
{\boldsymbol \Gamma}_{1\epsilon}&=&\frac{m_B}{\rho_\epsilon^2}[(1-2\omega_\epsilon^2){\mathbf A}_\epsilon -\epsilon (1+2\omega_\epsilon^2 +4A_\epsilon \omega_\epsilon){\mathbf v}_\epsilon] \,.\qquad
\eea

When we derived above the 1PM solution we only needed the leading-order (straight line) value of ${\boldsymbol \Gamma}_{1\epsilon}$, namely 
\bea
\Gamma_{1\epsilon}^x(\tau_A)&=& \frac{m_B(1-2\gamma^2)}{D^3(\tau_A)} b +O(G) \,,\nonumber\\
\Gamma_{1\epsilon}^y(\tau_A)
&=& \frac{m_B}{D^3(\tau_A)}\sqrt{\gamma^2-1} (2\gamma^2-3)\, \gamma \tau_A +O(G) ,\qquad
\eea
with the $x$ and $y$ components referring to the frame introduced at the end of Section II.
Note that these first-order expressions are independent of $\epsilon$, i.e. they are the same for the retarded
and advanced cases. Indeed, it is well known that radiation-reaction effects starts at order $G^2$, though
they show up only in a $O(G^2)$ loss of angular momentum \cite{Damour:1981bh,Damour:2020tta,Bini:2022wrq}.

The expression of $\Gamma_{2\epsilon}^\alpha[X_\epsilon]$ given in Ref. \cite{Bel:1981be} reads
\beq
\Gamma_{2\epsilon}^\alpha=\widetilde \Gamma_{2\epsilon}^\alpha-\epsilon \frac{d}{d\tau_A}[4m_A \Gamma_{1\epsilon}^\alpha \ln A_{\epsilon}]\,,
\eeq
where
\bea
\Gamma_{1\epsilon}^\alpha&=& \frac{m_B}{\rho_\epsilon^2}[(1-2\omega_\epsilon^2)A_\epsilon^\alpha -\epsilon(1+2\omega_\epsilon^2+4\omega_\epsilon A_\epsilon)v_\epsilon^\alpha]\,,\nonumber\\
\widetilde \Gamma_{2\epsilon}^\alpha&=& \frac{m_B}{\rho_\epsilon^3} [(m_A a +m_B a_0)A_\epsilon^\alpha +\epsilon(m_A c +m_B c_0)v_\epsilon^\alpha]\,,\nonumber\\
\eea
with
\bea
a_0&=& 2 [2\omega_\epsilon^2+(\omega_\epsilon+A_\epsilon)^2]\,,\nonumber\\
c_0&=&-2 [2\omega_\epsilon^2 +A_\epsilon(\omega_\epsilon+A_\epsilon)]\,,\nonumber\\
a&=& -\frac{2}{A_\epsilon^5}-5\frac{\omega_\epsilon}{A_\epsilon^4}+5 \frac{(1-2\omega_\epsilon^2)}{A_\epsilon^3}+2\omega_\epsilon \frac{(3-4\omega_\epsilon^2)}{A_\epsilon^2}\nonumber\\
&+&4 \frac{(2\omega_\epsilon^4+2\omega_\epsilon^2-1)}{A_\epsilon}+20\omega_\epsilon (2\omega_\epsilon^2-1)\nonumber\\
&+&12(2\omega_\epsilon^2-1) A_\epsilon\,,\nonumber\\
c&=& -\frac{47}{3 A_\epsilon^3}-32\frac{\omega_\epsilon}{A_\epsilon^2}+\frac{(3+16\omega_\epsilon^2-4\omega_\epsilon^4)}{A_\epsilon}\nonumber\\
&+& 20\omega_\epsilon (2\omega_\epsilon^2+3)+4 (3+26\omega_\epsilon^2) A_\epsilon+48\omega_\epsilon A_\epsilon^2\,,\qquad
\eea
where $a_0$, $c_0$, $a$, $c$  can be evaluated at the zeroth-order in $G$ (i.e., for example, with $\omega=-\gamma$ and ${\mathbf v}_\epsilon$ constant).

Explicit results for the various quantities $X_\epsilon=X_\epsilon^0+G X_\epsilon^1$ as functions of $\tau_A$ are listed in Table \ref{tab:scal} below.
All the first order scalar and vectorial correction terms have a mass structure, in the sense that they read
\beq
X_\epsilon^1=m_A X_{m_A\epsilon}^1+m_B X_{m_B\epsilon}^1\,.
\eeq

\begin{table*}  
\caption{\label{tab:scal}  List of the advanced/retarded corrections to the various quantities entering the acceleration of body A: $\tau_{B,\epsilon}$, $\omega_\epsilon$, $\rho_\epsilon$, ${\mathbf v}_\epsilon$, and ${\mathbf A}_\epsilon$ up to first order in $G$.
}
\begin{ruledtabular}
\begin{tabular}{ll}
$ \tau_{B,\epsilon}^0$ & $ \gamma \tau_A - \epsilon D(\tau_A)$\\
$ \tau_{B,\epsilon}^1$ & $ \frac{\epsilon}{\sqrt{\gamma^2-1}D(\tau_A)}\left[m_A\left((2\gamma^2-3)\gamma \tau_A \ln(S(\tau_{B,\epsilon}^0)) +\frac{(2\gamma^2-1) b S(\tau_{B,\epsilon}^0)}{\sqrt{\gamma^2-1}}\right)+ m_B\left((2\gamma^2-3)\gamma \tau_{B,\epsilon}^0 \ln(S(\tau_A)) +\frac{(2\gamma^2-1) b S(\tau_A)}{\sqrt{\gamma^2-1}}\right)  \right]$\\
\hline
$ \omega_\epsilon^0$ & $ -\gamma$\\
$ \omega_\epsilon^1$ & $ \frac{(2\gamma^2-3)\gamma}{D(\tau_A)}\left( \frac{m_A}{ A_\epsilon^0} +   m_B \right)$\\
\hline
$ \rho_\epsilon^0$ & $  D(\tau_A)$\\
$ \rho_\epsilon^1$ & $ -\epsilon \left[
\tau_{B,\epsilon}^1
+\frac{1}{D(\tau_{B,\epsilon}^0)}\left(\frac{(2\gamma^2-1)}{\sqrt{\gamma^2-1}}bS(\tau_{B,\epsilon}^0)+\gamma(2\gamma^2-3) \tau_A \right)m_A +\frac{\gamma(2\gamma^2-3)}{\sqrt{\gamma^2-1}}\ln(S(\tau_A)) m_B \right]$\\
\hline
${\mathbf v}_\epsilon^0$ & $u_{B\perp A}$\\
${\mathbf v}_\epsilon^1$ & $\frac{m_B \gamma (2\gamma^3-3)}{D(\tau_A)}u_A+\frac{2\gamma^2-1}{\sqrt{\gamma^2-1}}\left[m_B\gamma \frac{S(\tau_A)}{D(\tau_A)}+m_A \frac{S(\tau_{B,\epsilon}^0)}{D(\tau_{B,\epsilon}^0)}  \right] \hat b_{AB}- \frac{\gamma^2(2\gamma^2-3)}{\gamma^2-1}\left[\frac{m_A}{D(\tau_{B,\epsilon}^0)} +\frac{m_B}{D(\tau_A)} \right]u_{B\perp A}$\\
\hline
${\mathbf A}_\epsilon^0$ & $\frac{1}{D(\tau_A)}\left(b\hat b_{AB}-\tau_{B,\epsilon}^0 u_{B\perp A}  \right)$\\
${\mathbf A}_\epsilon^1$ & $m_B {\mathcal A}_1^{m_B} u_A +(m_A {\mathcal A}_2^{m_A}+m_B {\mathcal A}_2^{m_B}) \hat b_{AB}+(m_A {\mathcal A}_3^{m_A}+m_B {\mathcal A}_3^{m_B}) u_{B\perp A}$\\
${\mathcal A}_1^{m_B}$ & $ -\frac{1}{D^2(\tau_A)}\left[ \frac{b(2\gamma^2-1)}{\sqrt{\gamma^2-1}}S(\tau_A)+\gamma (2\gamma^2-3)\tau_{B,\epsilon}^0 \right] $\\
${\mathcal A}_2^{m_A} $ & $(2\gamma^2-1)\frac{S(\tau_{B,\epsilon}^0)}{D^2(\tau_A)}\left[
\frac{b^2\epsilon}{\sqrt{\gamma^2 - 1}D(\tau_{B,\epsilon}^0)} - \frac{\tau_A^2}{D(\tau_A)}
\right]+(2\gamma^2-3)\frac{b\gamma\tau_A}{D^2(\tau_A)}\left[
\frac{\epsilon}{D(\tau_{B,\epsilon}^0)} +  \frac{\ln(S(\tau_{B,\epsilon}^0))}{\sqrt{\gamma^2 - 1}D(\tau_A)}
\right]$\\
${\mathcal A}_2^{m_B}$ & $(2\gamma^2-1)\frac{S(\tau_A)}{D^2(\tau_A)}\left[\frac{\epsilon D(\tau_{B,\epsilon}^0)}{\sqrt{\gamma^2 - 1}} - \frac{\tau_A^2}{D(\tau_A)}
\right]+(2\gamma^2-3)\frac{b\gamma^2\tau_A\ln(S(\tau_A))}{\sqrt{\gamma^2 - 1}D^3(\tau_A)}$\\
${\mathcal A}_3^{m_A}$ & $-\frac{(2\gamma^2-1)S(\tau_{B,\epsilon}^0)b}{D^2(\tau_A)D(\tau_{B,\epsilon}^0)\sqrt{\gamma^2 - 1}}\left[
\epsilon\tau_{B,\epsilon}^0 + \frac{D(\tau_{B,\epsilon}^0)\gamma\tau_A}{\sqrt{\gamma^2 - 1}D(\tau_A)}
\right]+\frac{\gamma(2\gamma^2-3)}{D^2(\tau_A)}\left[\frac{b^2\gamma\ln(S(\tau_{B,\epsilon}^0))}{D(\tau_A)(\gamma^2 - 1)^{3/2}}
-\frac{\tau_A\epsilon\tau_{B,\epsilon}^0}{D(\tau_{B,\epsilon}^0)}
\right]$\\
${\mathcal A}_3^{m_B}$ & $-(2\gamma^2-1)\frac{b\gamma\tau_AS(\tau_A)}{(\gamma^2 - 1)D^3(\tau_A)}+\frac{\gamma(2\gamma^2-3)}{(\gamma^2-1)D^2(\tau_A)}\left[\frac{\gamma^2b^2\ln(S(\tau_A))}{\sqrt{\gamma^2 - 1}D(\tau_A)}-\epsilon D(\tau_{B,\epsilon}^0)
\right]$\\
\end{tabular}
\end{ruledtabular}
\end{table*}

After having obtained an explicit $G$-expansion of the acceleration of body as a function of $\tau_A$,
of the form of Eq. \eq{accexpl0}, the next step is to decompose 
$A_A^{\alpha}(\tau_A)= G A_A^{1\alpha}(\tau_A)+G^2 A_{A\,\epsilon}^{2\alpha}(\tau_A) +O(G^3)$
along our vectorial basis, i.e.
\bea
A_A^{\alpha}(\tau_A)&=& A_{u_A}(\tau_A) u_{A-}^\alpha+A_{\hat b_{AB}}(\tau_A) \hat b_{AB}^\alpha\nonumber\\
&+& A_{B\perp A}(\tau_A)u_{B\perp A}^\alpha\,.
\eea
The corresponding basis coefficients for
\bea
u_A^{\alpha}(\tau_A)&=& C_{u_A}(\tau_A) u_{A-}^\alpha+C_{\hat b_{AB}}(\tau_A) \hat b_{AB}^\alpha\nonumber\\
&+& C_{B\perp A}(\tau_A)u_{B\perp A}^\alpha
\eea
are then obtained by integration from $\tau_A= -\infty$, namely (remembering that we do not need to integrate
for $ C_{u_A}(\tau_A)$ in view of Eq. \eqref{eq_norm_u})
\bea
C_{\hat b_{AB}}(\tau_A) &=& \int_{-\infty}^{\tau_A} d \tau'_A A_{u_A}(\tau'_A)\,,\nonumber\\
C_{B\perp A}(\tau_A) &=& \int_{-\infty}^{\tau_A} d \tau'_A  A_{B\perp A}(\tau'_A)\,.
\eea
As discussed in \cite{Bini:2022wrq}, at the 2PM order, one can decompose the accelerations $A_A^{\alpha}(\tau_A)$
in a conservative part and a radiation-reaction one. As the 1PM contribution $ G A_A^{1\alpha}(\tau_A)$ is automatically
time-symmetric (and conservative), it is enough to decompose the 2PM contribution $ G^2 A_A^{2\alpha}(\tau_A)$ in
  conservative and dissipative parts: the conservative part being the time-even, half sum of the $\epsilon=+1$ and $\epsilon=-1$ values of $A_{A\,\epsilon}^{2\alpha}$, and the radiation-reaction part being the time-odd, half difference of the $\epsilon=+1$ and $\epsilon=-1$ values of $A_{A\,\epsilon}^{2\alpha}$:
\bea
A_{A\rm cons}^{2 \alpha}(\tau_A)&=&\frac12 \left(A_{A\,\epsilon=1}^{2\alpha}(\tau_A)+A_{A\,\epsilon=-1}^{2\alpha}(\tau_A)  \right)\,,\nonumber\\
A_{A\rm rr}^{2 \alpha}(\tau_A)&=& \frac12 \left(A_{A\,\epsilon=1}^{2\alpha}(\tau_A)-A_{A\,\epsilon=-1}^{2\alpha}(\tau_A) \right)\,.\qquad
\eea
Correspondingly, the integration of the equations of motion determines both a 2PM-accurate conservative solution, say
\bea
u_{A\rm cons}^\alpha(\tau_A)\,,\qquad z_{A\rm cons}(\tau_A)\,,
\eea
and an additional $O(G^2)$ radiation-reaction contribution, say
\bea
u_{A\rm rr}^\alpha(\tau_A)\,,\qquad z_{A\rm rr}(\tau_A)\,.
\eea
such that, e.g., the physical, retarded solution reads (at our 2PM order)
\bea
u_{A\rm ret}^\alpha(\tau_A) &=& u_{A\rm cons}^\alpha(\tau_A) + u_{A\rm rr}^\alpha(\tau_A) \,,\nonumber\\
 z_{A\rm ret}^\alpha(\tau_A) &=&
z_{A\rm cons}^\alpha(\tau_A) + z_{A\rm rr}^\alpha(\tau_A)\,.
\eea

We give below (separately for the conservative and radiation-reaction parts)  the explicit values of both 
the second-order corrections to the  velocity coefficients $C_{u_A}(\tau_A)$, $C_{\hat b_{AB}}(\tau_A)$ and $C_{B\perp A}(\tau_A)$ entering Eq. \eqref{u_A_con_Cs}, and the corresponding worldline basis coefficients.  
At order $G^2$, we find that all the second-order velocity coefficients  $C^{G^2}_{u_A}(\tau_A)$, $C^{G^2}_{\hat b_{AB}}(\tau_A)$ and $C^{G^2}_{B\perp A}(\tau_A)$ decay at least like $O(\frac1{\tau_A^2})$ when $\tau_A \to - \infty$. 
On the other hand, when 
$\tau_A \to + \infty$,  the conservative parts of $C^{G^2}_{u_A}(\tau_A)$, $C^{G^2}_{\hat b_{AB}}(\tau_A)$ and $C^{G^2}_{B\perp A}(\tau_A)$ tend to non-zero constants, while their radiation-reaction parts all vanish at least like $O(\frac1{\tau_A^2})$; more precisely
$C^{G^2}_{\hat b_{AB} \, \rm rr}(\tau_A)\sim \frac{1}{\tau_A^3}$ and $C^{G^2}_{B\perp A \, \rm rr}(\tau_A)\sim \frac{1}{\tau_A^2}$.

As a consequence, the corresponding orbit coefficients are simply given 
by integrating them between $ - \infty$ and $\tau_A$:
\bea
z_{A2}^\alpha&=&\int_{-\infty}^{\tau_A} d\tau_A u_{A2}^\alpha(\tau_A)\,.
\eea
We recall that at first order in $G$ we could not write a similar expression for $z_{A1}^\alpha$ due to the logarithmic divergence of its $u_{B\perp A}$-component. 

We have given in Eq. \eqref{intCG1} above the explicit values of the (dimensionless) integrals entering the 1PM orbit,
namely
\bea
{\mathcal I}^{G^1 m_B}_{\hat b_{AB}}(\tau_A)&=&\int^{\tau_A}_{-\infty} C^{G^1 m_B}_{\hat b_{AB}}(\xi) d\xi=\frac{1-2\gamma^2}{\gamma^2-1}S(\tau_A)
\,,\nonumber\\
{\mathcal I}^{G^1 m_B}_{B\perp A}(\tau_A)&=&\int_{0}^{\tau_A} C^{G^1 m_B}_{B\perp A}(\xi) d\xi=\frac{\gamma (2\gamma^2-3)}{(\gamma^2-1)^{3/2}} \ln S(\tau_A)
\, .\nonumber\\
\eea

At the 2PM level, we needed to  evaluate the following integrals (which have the dimension of an inverse length)
\bea \label{intG2}
{\mathcal I}^{G^2 m_B^2}_{u_A}(\tau_A)&=&\int^{\tau_A}_{-\infty} C^{G^2 m_B^2}_{u_A}(\xi) d\xi
\,,\nonumber\\
{\mathcal I}^{G^2 m_A m_B}_{\hat b_{AB}}(\tau_A)&=&\int^{\tau_A}_{-\infty} C^{G^2 m_A m_B}_{\hat b_{AB}}(\xi) d\xi
\,,\nonumber\\
{\mathcal I}^{G^2 m_B^2}_{\hat b_{AB}}(\tau_A)&=&\int^{\tau_A}_{-\infty} C^{G^2 m_B^2}_{\hat b_{AB}}(\xi) d\xi
\,,\nonumber\\
{\mathcal I}^{G^2 m_A m_B}_{B\perp A}(\tau_A)&=&\int^{\tau_A}_{-\infty} C^{G^2 m_A m_B}_{B\perp A}(\xi) d\xi
\,,\nonumber\\
{\mathcal I}^{G^2 m_B^2}_{B\perp A}(\tau_A)&=&\int^{\tau_A}_{-\infty} C^{G^2 m_B^2}_{B\perp A}(\xi) d\xi
\,,
\eea
in terms of which the parametric equation of the orbit of the body A reads
\bea
\label{orbittauA}
&&z_{A}^\alpha(\tau_A)=b_A^\alpha +u_{A-}^\alpha \tau_A\nonumber\\
&&\quad +
 Gm_B \left[ \hat b_{AB}^\alpha \, {\mathcal I}^{G^1 m_B}_{\hat b_{AB}}(\tau_A) + u_{B\perp A}^\alpha \, {\mathcal I}^{G^1 m_B}_{B\perp A}(\tau_A) \right]\nonumber\\
&&\quad +   G^2  \left[ 
u_{A-}^\alpha\,  m_B^2 \, {\mathcal I}^{G^2 m_B^2}_{u_A}(\tau_A) \right.\nonumber\\
&&\quad +
\hat b_{AB}^\alpha \left(m_A m_B \, {\mathcal I}^{G^2 m_A m_B}_{\hat b_{AB}}(\tau_A)+ m_B^2 \, {\mathcal I}^{G^2 m_B^2}_{\hat b_{AB}}(\tau_A)\right)\nonumber\\
&&\quad + \left. u_{B\perp A}^\alpha \left(m_A m_B {\mathcal I}^{G^2 m_A m_B}_{B\perp A}(\tau_A) + m_B^2\,{\mathcal I}^{G^2 m_B^2}_{B\perp A}(\tau_A)\right) \right]\,.\nonumber\\
\eea

As above, the orbit of  body $B$ is simply obtained by exchanging the labels $A$ and $B$ (remembering that $\hat b_{BA}=-\hat b_{AB}$).

We express  all coefficients (for body $A$) as functions of the dimensionless parameter 
\beq \label{Tvstau}
T  \equiv{\sqrt{\gamma^2-1}}{b} \tau_A\,,
\eeq
in terms of which the basic quantities \eqref{D_and_S_defs} read
\beq
D(T)= b\sqrt{1+T^2}\,,\qquad
S(T)= T+\sqrt{1+T^2}\,.
\eeq

\section{The conservative solution at $O(G^2)$}

The $O(G^2)$ conservative part of the four velocity of the body A is given by Eq. \eqref{u_A_con_Cs} with coefficients 
(expressed in terms of $T$, via Eq. \eq{Tvstau})
\bea
C_{u_A\,{\rm cons}}^{G^2}(\tau_A) &=&\frac{m_B^2}{b^2}C_{u_A\,{\rm cons}}^{G^2m_B^2}(T)
\,, \nonumber\\
C_{\hat b_{AB}\,{\rm cons}}^{G^2}(\tau_A)&=&\frac{m_B m_A}{b^2} C_{\hat b_{AB}\,{\rm cons}}^{G^2m_Am_B}(T) +\frac{m_B^2}{b^2}C_{\hat b_{AB}\,{\rm cons}}^{G^2m_B^2}(T)
\,, \nonumber\\
C_{B\perp A\,{\rm cons}}^{G^2}(\tau_A)&=& \frac{m_B m_A}{b^2} C_{B\perp A\,{\rm cons}}^{G^2 m_A m_B}(T)+\frac{m_B^2}{b^2} C_{B\perp A\,{\rm cons}}^{G^2m_B^2}(T) \,,\nonumber\\
\eea  
with
\begin{widetext}
\bea
C_{u_A\,{\rm cons}}^{G^2m_B^2}(T)&=& \frac{\gamma^2 (2\gamma^2 - 3 )^2}{2 (T^2 + 1) (\gamma^2 - 1)} 
+\frac{ (T+\sqrt{1+T^2})^2 (2\gamma^2 -1)^2}{2(T^2 + 1) (\gamma^2 - 1)  } \,,\nonumber\\
C_{\hat b_{AB}\,{\rm cons}}^{G^2m_Am_B}(T)&=&(2\gamma^2-1)F(T,\gamma)-\frac{3(5\gamma^2-1)}{4(\gamma^2-1)^{1/2}}  \left({\rm arctan}\left( \frac{T}{\gamma}\right) +\frac{\pi}{2}\right)\nonumber\\
&+& \frac{\gamma T}{\sqrt{\gamma^2-1}(T^2+\gamma^2)} \left[\frac{17}{4}\gamma^2-\frac{13}{4}+\frac{\left(\frac{49}{6}\gamma^2-\frac{20}{3} \right)(\gamma^2-1)}{T^2+\gamma^2}  
+\frac{\left(-\frac{14}{3}\gamma^2+2 \right)(\gamma^2-1)^{2}}{(T^2+\gamma^2)^2} 
+\frac{4\gamma^2(\gamma^2-1)^{3}}{(T^2+\gamma^2)^3} 
 \right]\nonumber\\
&+&\frac{\gamma (2\gamma^2-1)}{(\gamma^2-1)^{3/2} \sqrt{1+T^2}}\left[2\gamma^2-3
-\frac{(2\gamma^2-1) T}{\sqrt{1+T^2}}
+\frac{2\gamma^2-1}{1+T^2}\right]
\,,\nonumber\\
C_{\hat b_{AB}\,{\rm cons}}^{G^2m_B^2}(T)&=&  \frac{\gamma^2(2\gamma^2-3) (2\gamma^2-1)}{(\gamma^2-1)^{3/2}\sqrt{1+T^2}}\left(\frac{{\rm arcsinh}(T)}{1+T^2}+1\right) -\frac{3(5\gamma^2-1)}{4(\gamma^2-1)^{1/2} }\left({\rm arctan}(T)+\frac{\pi}{2}\right)\nonumber\\
&+&\frac{T (16\gamma^6-47\gamma^4+30\gamma^2-3)}{4(\gamma^2-1)^{3/2}(T^2+1)}
+\frac{(2\gamma^2-1)^2}{(\gamma^2-1)^{3/2}(T^2+1)^{3/2}}
-\frac{T (\gamma^2-1)^{1/2}  }{2 (T^2+1)^2} 
\,,\nonumber\\
C_{B\perp A\,{\rm cons}}^{G^2 m_A m_B}(T)&=&-\frac{\gamma (2\gamma^2-3)}{(\gamma^2-1)^{1/2}}TF(T,\gamma)
+\frac{2\gamma^2-1}{(\gamma^2-1)^2}\left[2\gamma^2-1
+\frac{\gamma^2(2\gamma^2-3)}{(T^2+1)}\right]\left(1+\frac{T}{\sqrt{1+T^2}}\right)\nonumber\\
&-&\frac{(\frac23\gamma^4+\frac83\gamma^2-\frac{13}{3})}{(\gamma^2-1)(T^2+\gamma^2)} 
+\frac{(-\frac{14}{3}\gamma^4+\frac{17}{3}\gamma^2+\frac12)}{(T^2+\gamma^2)^2}
-\frac{4(\gamma^2-1)\gamma^2 (5\gamma^2-3)}{3(T^2+\gamma^2)^3}  
+\frac{4 (\gamma^2-1)^2\gamma^4}{(T^2+\gamma^2)^4}
\,,\nonumber\\
C_{B\perp A\,{\rm cons}}^{G^2m_B^2}(T) &=& \frac{\gamma T}{(\gamma^2-1)^2\sqrt{1+T^2}}\left[(2\gamma^2-3)\frac{\gamma^2(2\gamma^2-3){\rm arcsinh}(T)+2\gamma^2-1}{1+T^2}
+(2\gamma^2-1)^2\right]\nonumber\\
&+&\frac{\gamma (2\gamma^2-1)^2}{(\gamma^2-1)^2}
-\frac{\gamma (4\gamma^6-16\gamma^4+20\gamma^2-7)}{(\gamma^2-1)^2(T^2+1)}
-\frac{\gamma}{2(T^2+1)^2}
 \,,
\eea
where
\beq
\label{F_def}
F(T,\gamma)=\frac{1}{(1+T^2)^{3/2}}\left[ -4{\rm arctanh}\left( \frac{T\sqrt{\gamma^2-1}}{\gamma\sqrt{1+T^2}}\right)+\frac{\gamma (2\gamma^2-3) }{(\gamma^2-1)^{3/2}} {\rm arcsinh}(T) \right]\,.
\eeq
\end{widetext} 
A companion ancillary file associated with the arXiv version of the paper contains all these explicit expressions.

When $T\to -\infty$ we find
\bea
C_{\hat b_{AB}\,{\rm cons}}^{G^2m_Am_B}(T)\,,\quad C_{\hat b_{AB}\,{\rm cons}}^{G^2m_B^2}(T)&\sim & O\left( \frac{1}{T^3}\right)\,, \nonumber\\
C_{B\perp A\,{\rm cons}}^{G^2 m_A m_B}(T)\,,\quad C_{B\perp A\,{\rm cons}}^{G^2m_B^2}(T)&\sim&  O\left( \frac{1}{T^2}\right)\,.
\eea
The corresponding values of the integrals \eq{intG2}  entering the 2PM orbit \eq{orbittauA} are given in
Table \ref{tab:calIcons} below.

\begin{table*}  
\caption{\label{tab:calIcons} 
All the integrals entering the solution for the conservative orbit are listed below using the dimensionless temporal variable $T$, see  Eq. \eq{Tvstau}.
We use the notation   $Z(T,\gamma)= \frac{\gamma^2-1}{T^2+\gamma^2}$ and $F(T,\gamma)=\frac{1}{(1+T^2)^{3/2}}\left[ -4{\rm arctanh}\left( \frac{T\sqrt{\gamma^2-1}}{\gamma\sqrt{1+T^2}}\right)+\frac{\gamma (2\gamma^2-3) }{(\gamma^2-1)^{3/2}} {\rm arcsinh}(T) \right]$, see Eq. \eqref{F_def}.}
\begin{ruledtabular}
\begin{tabular}{ll}
$b\,{\mathcal I}^{G^2 m_B^2}_{u_A}(T)$&
$\frac{(4\gamma^4-12\gamma^2+1)}{2\sqrt{\gamma^2-1}}\left({\arctan (T)+\frac{\pi}{2}}\right)+\frac{ (2\gamma^2-1)^2 }{(\gamma^2-1)^{3/2}}[T+\sqrt{T^2+1}]$
\\
$b\,{\mathcal I}^{G^2 m_A m_B}_{\hat b_{AB}}(T)$&
$ -\frac{3 T (5\gamma^2-1)}{4(\gamma^2-1)} \left({\arctan \left(\frac{T}{\gamma}\right)+\frac{\pi}{2}}\right)+\frac{(T^2+1) (2\gamma^2-1) T}{\sqrt{\gamma^2-1}} F
-\frac{2\gamma^3 Z^3}{ 3 (\gamma^2-1)}
+\frac{\gamma (7\gamma^2-3)Z^2}{6(\gamma^2-1)} -\frac{\gamma (49\gamma^2-40) Z}{12 (\gamma^2-1)}$\\
&$+\frac{(2\gamma^2-3)\gamma (2\gamma^2-1)}{(\gamma^2-1)^2} {\rm arcsinh}(T)+\frac{\gamma (\gamma^2+1)^2}{4 (\gamma^2-1)^2} +\frac{4(2\gamma^2-1)}{\sqrt{\gamma^2-1}}{\rm arccosh}(\gamma)  
+\frac{(2\gamma^2-1)^2\gamma T}{ (\gamma^2-1)^2\sqrt{T^2+1} }$
\\
$b\, {\mathcal I}^{G^2 m_B^2}_{\hat b_{AB}}(T)$&
$-\frac{3 T (5\gamma^2-1)}{4(\gamma^2-1) }\left({\arctan (T)+\frac{\pi}{2}}\right)+{\rm arcsinh}(T)\frac{[T+\sqrt{T^2+1}](2\gamma^2-1) (2\gamma^2-3)\gamma^2}{\sqrt{T^2+1}(\gamma^2-1)^2}
+\frac{ (\gamma^2+1)^2}{4(\gamma^2-1)^2}+\frac{1}{ 4(1+T^2)} +\frac{(2\gamma^2-1)^2 T}{(\gamma^2-1)^2\sqrt{T^2+1}} $
\\
$b\, {\mathcal I}^{G^2 m_A m_B}_{B\perp A}(T)$&
$ \frac{3\gamma (5\gamma^2-9) }{ 4 (\gamma^2-1)^{3/2}}\left({\arctan \left(\frac{T}{\gamma}\right)+\frac{\pi}{2}}\right)-\frac{\gamma (2\gamma^2-3) (T^2+1) }{(\gamma^2-1)} F+
\frac{2 T\gamma ^2 Z^3}{3 (\gamma^2-1)^{3/2}}-\frac{ T (5\gamma^2-1) Z^2}{ 6 (\gamma^2-1)^{3/2}} 
-\frac{ T (43\gamma^2-52) Z}{12 (\gamma^2-1)^{3/2}}$\\
&$-\frac{\gamma^2 (2\gamma^2-1) (2\gamma^2-3)}{ (\gamma^2-1)^{5/2}\sqrt{T^2+1}}+\frac{(2\gamma^2-1)^2 [T+\sqrt{T^2+1}]}{ (\gamma^2-1)^{5/2}}$
\\
$b\, {\mathcal I}^{G^2 m_B^2}_{B\perp A}(T)$&
$ \frac{3\gamma (5\gamma^2-9) }{ 4 (\gamma^2-1)^{3/2}}\left({\arctan (T)+\frac{\pi}{2}}\right)
-\frac{(2\gamma^2-3)^2\gamma^3}{(\gamma^2-1)^{5/2}\sqrt{T^2+1}}{\rm arcsinh}(T)-\frac{(2\gamma^2-1) (2\gamma^2-3)\gamma}{(\gamma^2-1)^{5/2}\sqrt{T^2+1}} 
+\frac{(2\gamma^2-1)^2\gamma [T+\sqrt{T^2+1}]}{(\gamma^2-1)^{5/2}}$\\
&$-\frac{\gamma T}{ 4\sqrt{\gamma^2-1}(T^2+1) } $
\\
\end{tabular}
\end{ruledtabular}
\end{table*}

\section{The $O(G^2)$ radiation-reaction contribution}

The $O(G^2)$ radiation-reaction contribution to the four velocity of the body A is given by Eq. \eqref{u_A_con_Cs} with,
notably,
\be
C_{u_A\,{\rm rr}}^{G^2}(\tau_A) =0
\ee
and with nonvanishing coefficients
\bea
C_{\hat b_{AB}\,{\rm rr}}^{G^2}(\tau_A)&=&\frac{m_B m_A}{b^2} C_{\hat b_{AB}\,{\rm rr}}^{G^2m_Am_B}(T) \,, \nonumber\\
C_{B\perp A\,{\rm rr}}^{G^2}(\tau_A)&=& \frac{m_B m_A}{b^2} C_{B\perp A\,{\rm rr}}^{G^2 m_A m_B}(T) \,,\nonumber\\
\eea  
with 
\begin{widetext}
\bea
C_{\hat b_{AB}\,{\rm rr}}^{G^2m_Am_B}(T)&=& (2\gamma^2-1)H(T,\gamma)
-\frac{3 (5\gamma^2-1)}{4(\gamma^2-1)^{1/2}} \left[{\rm arctan}\left(\frac{\sqrt{1+T^2}}{\sqrt{\gamma^2-1}}\right)-\frac{\pi}{2}\right]   \nonumber\\ 
&+& \frac{Z(T,\gamma)}{(T^2+1)^{3/2}}\left[-\frac{67\gamma^4-74\gamma^2+15}{4} 
+ \frac{ (43\gamma^4-30\gamma^2+1)Z(T,\gamma)}{2}\right.\nonumber\\
&-&\left. 
\frac{2\gamma^2 (19\gamma^2-6)Z^2(T,\gamma)}{3 }  
+  4\gamma^4 Z^3(T,\gamma)  \right] \nonumber\\
&+& -\frac{3(5\gamma^2-1)}{4(T^2+1)^{1/2}} 
+\frac{(-3+47\gamma^6+33\gamma^2-89\gamma^4)}{12(\gamma^2-1)(T^2+1)^{3/2}} 
\,,\nonumber\\
C_{B\perp A\,{\rm rr}}^{G^2 m_A m_B}(T) &=&\frac{(2\gamma^2-3)\gamma T}{\sqrt{\gamma^2-1}}\left(H(T,\gamma)+\frac{5\gamma^2-8}{3(1+T^2)^{3/2}}\right)
-\frac{\gamma T Z(T,\gamma)}{ \sqrt{\gamma^2-1}(1+T^2)^{3/2}}\left[ 2 (2\gamma^2-3) 
\right.\nonumber\\
&+&\left.  Z(T,\gamma)(2\gamma^2+5)  
-\frac{2 Z^{2}(T,\gamma) (16\gamma^2-3) }{ 3  }
+ 4 Z^{3}(T,\gamma)  \gamma^2 
\right]
  \,,\nonumber\\
\eea  
where
\bea
\label{H_def}
H(T,\gamma)&=& \frac{1}{(T^2+1)^{3/2}}\left[2\ln \left(\frac{T^2+\gamma^2}{T^2+1} \right)
-\frac{\gamma (2\gamma^2-3) }{ (\gamma^2-1)^{3/2}}{\rm arccosh}(\gamma)\right]\,,\nonumber\\
Z(T,\gamma)&=& \frac{\gamma^2-1}{T^2+\gamma^2}\,.
\eea
\end{widetext}
In the limit $T\to  \mp\infty$ we find 
\bea  \label{decayrr}
C_{\hat b_{AB}\,{\rm rr}}^{G^2m_Am_B}(T)&\sim& O\left( \frac{1}{T^3} \right)\,,\nonumber\\
C_{B\perp A\,{\rm rr}}^{G^2 m_A m_B}(T) &\sim& O\left( \frac{1}{T^2} \right)\,.
\eea

The corresponding radiation-reaction contributions to the integrals \eq{intG2} entering the 2PM orbit \eq{orbittauA} are given in
Table \ref{tab:calIrr} below.

\begin{table*}  
\caption{\label{tab:calIrr} 
All the integrals entering the solution for the radiation-reaction orbit are listed below as in Table \ref{tab:calIrr}. We use the notation $Z(T,\gamma)= \frac{\gamma^2-1}{T^2+\gamma^2}$ and $H(T,\gamma)= \frac{1}{(T^2+1)^{3/2}}\left[2\ln \left(\frac{T^2+\gamma^2}{T^2+1} \right)
-\frac{\gamma (2\gamma^2-3) }{ (\gamma^2-1)^{3/2}}{\rm arccosh}(\gamma)\right]$, see Eq. \eqref{H_def}.}
\begin{ruledtabular}
\begin{tabular}{ll}
$b\, {\mathcal I}^{G^2 m_A m_B}_{\hat b_{AB}}(T)$&
$  -\frac{2T\gamma^2 Z^3}{3\sqrt{1+T^2}\sqrt{\gamma^2-1}}  +\frac{ T (14\gamma^2-2) Z^2}{12 \sqrt{1+T^2}\sqrt{\gamma^2-1}} +
\frac{T (-49\gamma^2+37) Z}{ 12\sqrt{1+T^2}\sqrt{\gamma^2-1}}
-\frac{3 T (5\gamma^2-1)}{4 (\gamma0^2-1)}\left( {\rm arctan}\left(\sqrt{\frac{1+T^2}{\gamma^2-1}}\right)-\frac{\pi}{2}\right)$\\
&$+\frac{(2\gamma^2-1) T (1+T^2) H}{\sqrt{\gamma^2-1}}
-\frac{T(5\gamma^4+30\gamma^2-23)}{12\sqrt{1+T^2}(\gamma^2-1)^{3/2}} $
\\
$b\, {\mathcal I}^{G^2 m_A m_B}_{B\perp A}(T)$&
$ -\frac{2\gamma^3Z^3}{3(\gamma^2-1)\sqrt{1+T^2}}  
+\frac{\gamma (3\gamma^2-1) Z^2 }{2 (\gamma^2-1)\sqrt{1+T^2}}
+\frac{\gamma (11\gamma^2-15) Z}{4 (\gamma0^2-1)\sqrt{1+T^2}} 
-\frac{ (2\gamma^2-3)\gamma(1+T^2) H}{(\gamma^2-1)} 
+\frac{4 (2\gamma^2-3)\gamma\ln(\gamma)}{(\gamma^2-1)}-\frac{(2\gamma^2-3)^2\gamma^2{\rm arccosh}(\gamma)}{(\gamma^2-1)^{5/2}} $\\
&$+\frac{3(5\gamma^2-9)\gamma}{4 (\gamma^2-1)^{3/2}} \left( {\rm arctan}\left(\sqrt{\frac{1+T^2}{\gamma^2-1}}\right)-{\rm arctan}\left(\sqrt{\frac{1}{\gamma^2-1}}\right)  \right)
-\frac{(48\gamma^8-141\gamma^6+132\gamma^4-49\gamma^2-2)}{12\gamma^3 (\gamma^2-1)^2} 
+\frac{\gamma (5\gamma^4-2\gamma^2-15)}{12 (\gamma^2-1)^2\sqrt{1+T^2}}  $
\\
\end{tabular}
\end{ruledtabular}
\end{table*}

\section{Some consequences of the 2PM-accurate solution}

We have performed several checks of our results on the 2PM-accurate solution of the equations of motion.

First, in view of the fast decay, Eq. \eq{decayrr}, of the radiation-reaction contributions to $u_A(\tau_A)$ we see
that the total change in the linear momentum $p_A^\alpha(\tau_A)= m_A u_A^\alpha(\tau_A)$ of body $A$
(between $\tau_A= - \infty$ and $\tau_A= + \infty$) is entirely given by the conservative contribution to $u_A(\tau_A)$.

One then find that  the total variation 
$\Delta p_{A }^{\alpha}=\lim_{\tau_A\to \infty} p_{A}^{\alpha}-\lim_{\tau_A\to -\infty} p_{A}^{\alpha}$ of $p_A^\alpha(\tau_A)= m_A u_A^\alpha(\tau_A)$ during scattering reads
\bea
\Delta p_{A}^{\alpha}&=& \Delta p_{A \, \rm cons}^{\alpha} \nonumber\\
&=&   D_{u_A} u_{A-}^\alpha+D_{\hat b_{AB}} \hat b_{AB}^\alpha +
D_{B\perp A}u_{B\perp A}^\alpha\,,\qquad
\eea
with
\bea
 D_{u_A} &=&\frac{G^2 m_Am_B^2}{b^2}\frac{2(2\gamma^2-1)^2}{\gamma^2-1}  +O(G^3)
\,,\nonumber\\
 D_{\hat b_{AB}}&=& -\frac{G m_Am_B}{b}\frac{2(2\gamma^2-1)}{\sqrt{\gamma^2-1}}\nonumber\\
&-& \frac{G^2 m_Am_B}{b^2}(m_A+m_B)\frac{3\pi(5\gamma^2-1)}{4\sqrt{\gamma^2-1}}
 +O(G^3)
\,,\nonumber\\
 D_{B\perp A}&=&\frac{G^2 m_Am_B}{b^2}(m_A+m_B)\frac{2(2\gamma^2-1)^2}{\gamma^2-1}
 +O(G^3)\,.\nonumber\\
\eea

It is then easily checked that this result is equivalent to the 2PM-accurate value of the scattering angle
first derived by Westpfahl \cite{Westpfahl:1985tsl}, namely
\bea \label{chiexpj}
\frac12 \chi&=&  \chi_1 \frac{G m_A m_B}{J}+  \chi_2 \left( \frac{G m_A m_B}{J}\right)^2\nonumber\\ 
&+& O\left( \left( \frac{G m_A m_B}{J}\right)^3\right) \,,
\eea
with  a total {\it incoming} angular momentum [in the center-of-mass (cm) frame] given by
\be \label{J}
J = b \, P_{\rm cm}\,,
\ee
 where (see, e.g., Ref. \cite{Damour:2016gwp}):
$E_{\rm cm} P_{\rm cm} = m_A m_B \sqrt{\gamma^2-1}$
and  $E_{\rm cm}= (m_A+m_B) h$, with the notation $h \equiv \sqrt{1 + 2 \frac{m_A m_B}{(m_A+m_B)^2}(\g-1)}$.
The 1PM and 2PM scattering-angle expansion coefficients $\chi_1$ and $\chi_2$ entering Eq. \eq{chiexpj} read
\beq
\label{chi_12}
\chi_1=\frac{2\gamma^2-1}{\sqrt{\gamma^2-1}}\,,\qquad \chi_2=\frac{3 \pi}{8} \frac{5\gamma^2-1}{h}\,.
\eeq
Eq. \eq{J} shows that the  parameter $b$ that entered the 2PM-accurate solution for the worldlines given above is equal
to the usual (incoming) impact parameter. 

Though, as mentioned, the two-body system does not loose energy and
linear momentum at order $G^2$ during scattering, it does loose angular momentum. The radiation of angular momentum
at infinity through the emission of gravitational waves at order $G^2$ was first derived in Ref. \cite{Damour:2020tta}
(and first extended to the $G^3$ accuracy in Ref. \cite{Manohar:2022dea}). 
A direct proof that the $O(G^2)$ radiated loss of angular momentum is balanced by a loss of the ``mechanical" 
angular momentum of the two-body system was given in Ref. \cite{Bini:2022wrq}. We can use the explicit 
 2PM-accurate  worldline solution given above to provide an even more direct check of the loss of
 mechanical angular momentum of the two-body system during scattering. We did this check  in two slightly different
 ways: (i) by using the Fokker-Wheeler-Feynman definition of a  Poincar\'e-covariant ``conserved"\footnote{In this covariant
 approach $J^{\rm cons}_{\mu\nu}$ is the sum of the usual mechanical  $J_{\mu\nu}$ and of an interaction contribution,
 discussed at order $G^1$ in  \cite{Bini:2022wrq}. One can then use the fact that the $O(G^2)$ interaction contribution
 tends to zero in the asymptotic past and the asymptotic future.}
  $J^{\rm cons}_{\mu\nu}$
 associated to the conserved part of the 2PM dynamics; and (ii) by using the ordinary mechanical angular momentum ${\bf J}$
 in the cm frame. Let us present here the results of the second (technically simpler) approach.
 
 Let us work in the incoming cm frame, i.e. ${\mathbf p_{A-}} + {\mathbf p_{B-}}= {\boldsymbol 0}$, with coordinate time $t$, and consider the total mechanical angular momentum of the two-body system, i.e.
 \be \label{Jcm}
 {\mathbf J}^{\rm mech}(t)= {\mathbf z_A}(\tau_A(t)) \times {\mathbf p_A}(\tau_A(t)) + {\mathbf z_B}(\tau_B(t)) \times {\mathbf p_B}(\tau_B(t)).
 \ee
Here, $\tau_A(t)$ and $\tau_B(t)$ are the proper times on the two worldlines corresponding to the cm-frame time $t$,
i.e. such that $ z_A^0(\tau_A)=t=  z_B^0(\tau_B)$. 

Inserting in Eq. \eq{Jcm} the explicit $G^2$-accurate solutions for $z_A, z_B, u_A$,  and $u_B$ (and perturbatively solving
$ z_A^0(\tau_A)=t=  z_B^0(\tau_B)$) we find that the value of $ {\mathbf J}^{\rm mech}(t)$ in the asymptotic past
is equal to
\be
{\mathbf J}^{\rm mech}(t=-\infty)= b \,  P_{\rm cm}\,  e_z\,,
\ee
with $e_z$  orthogonal to the orbital plane (see end of Sec. \ref{setup}).
The total variation of mechanical cm angular momentum during scattering,
\be
\Delta  {\mathbf J}^{\rm mech}=  {\mathbf J}^{\rm mech}(t=+\infty)-  {\mathbf J}^{\rm mech}(t=-\infty)\,,
\ee
is then found to be equal to
\be
\Delta  {\mathbf J}^{\rm mech}=  -\frac{G^2m_A m_B }{b^2}\frac{2\gamma^2-1}{\sqrt{\gamma^2-1}}  {\mathcal I}(\gamma) \,{\mathbf b}_{AB}\times ({\mathbf  p_A} -  {\mathbf p_B} ) \,,
\ee
i.e. $\Delta  {\mathbf J}^{\rm mech}=  \Delta J_{\rm rr} \, e_z$ with
\be \label{Jrr}
\frac{\Delta J_{\rm rr}}{  b \, P_{\rm cm}}= -\frac{ G^2m_A m_B }{b^2}\frac{2(2\gamma^2-1)}{\sqrt{\gamma^2-1}}  {\mathcal I}(\gamma) \,,
\ee
where we denoted
\bea
{\mathcal I}(\gamma)&=&-\frac{16}{3}+\frac{2\gamma^2}{\gamma^2-1}\nonumber\\
&+& 4\frac{\gamma(2\gamma^2-3)}{(\gamma^2-1)^{3/2}}{\rm arcsinh}\left(\sqrt{\frac{\gamma-1}{2}}\right)\,.
\eea
The result \eq{Jrr} agrees with (minus) the  radiated fluxes derived in Refs. \cite{Damour:2020tta,Manohar:2022dea},
and with the radiation-reaction-force computation of Ref. \cite{Bini:2022wrq}.

  Refs. \cite{Iyer:1995rn,Nissanke:2004er,Blanchet:2018yqa} have derived the radiation-reaction part
 of the acceleration of body A in harmonic coordinates at the 3.5 PN level, namely
\bea \label{ArrPN}
{\mathbf A}_A^{\rm rr\, PN}&=&\frac{G^2 m_A m_B}{r_{12}^3}\left[{\mathcal A}_A^{\rm PN}{\mathbf n}_{12}
+ {\mathcal B}_A^{\rm PN}{\mathbf v}_{12}\right]\,,
\eea
where the coefficients
\bea
{\mathcal A}_A^{\rm PN}&=&{\mathcal A}_A^{2.5\rm PN}+{\mathcal A}_A^{3.5 \rm PN}+\ldots\,,\nonumber\\
{\mathcal B}_A^{\rm PN}&=&{\mathcal B}_A^{2.5 \rm PN}+{\mathcal B}_A^{3.5 \rm PN}+\ldots\,,
\eea
were obtained as functions of ${\mathbf z}_{12}(t)$,  ${\mathbf v}_{1}(t)$,  ${\mathbf v}_{2}(t)$ in a general
Lorentz frame. The latter coefficients have a polynomial structure in $G$. By keeping only the $G^0$ contributions
in them Eq. \eq{ArrPN} yields the 3.5PN-accurate value of the 
2PM piece of the radiation-reaction acceleration in a general Lorentz frame.
Higher-PN-order results (namely 4.5PN) on radiation-reaction accelerations have been obtained in Refs.  \cite{Gopakumar:1997ng,Leibovich:2023xpg,Blanchet:2024loi}. However, the latter results have not been derived
in  harmonic coordinates, preventing us from making any  direct comparison.

For simplicity, we checked our 2PM-accurate results on ${\mathbf A}_A^{\rm rr\, PM}$ in the incoming
 rest frame of body A (i.e. ${\mathbf u}_{A-}=0$).  As the radiation-reaction starts at order $G^2$ we can
 approximate the worldlines by straight lines. In the incoming rest frame of body $A$, the coordinate
 time is then $t=z^0_A$. The corresponding   proper time, $\tau_B(t)$, along the world line of body $B$, $\tau_B$, is related to $t$ by the condition $z_A^0(\tau_A)-z_B^0(\tau_B)=0$ which yields
\beq
\tau_B=\frac{\tau_A}{\gamma}+ O(G)=\frac{t}{\gamma} + O(G)\,.
\eeq
We then find, for instance, (henceforth labelling the two bodies as 1 and 2, rather than $A$ and $B$)
\bea
{\mathbf z}_1(\tau_A(t))-{\mathbf z}_2(\tau_B(t))=  b \, {e}_x - {\mathbf v}_2  t + O(G),
\eea
where $ {e}_x$ is the same direction (along the incoming impact parameter) as above.

When comparing our results with PN ones, we must also express $\g$ in terms of $v_2$, using (in the $A$-frame)
$\g = \frac1{\sqrt{1-v_2^2}}$.

Using the notation
\be \label{defDX}
D_{\rm PN}\equiv \sqrt{b^2+v_2^2t^2}=r_{12}(t) + O(G)\; ; \;  X \equiv\frac{v_2 t}{D_{\rm PN}}\,,
\ee
the beginning of the PN expansion of the  $x$ and $y$ components (in the notation of the
end of Sec. \ref{setup}) of our 2PM radiation-reaction acceleration 
reads
\begin{widetext}
\bea
\label{A2535}
{\mathbf A}^{2.5+3.5}_{x\, \rm PN\, from \, PM}&=&\frac{G^2 m_Am_B}{D_{\rm PN}^3} \frac{b}{D_{\rm PN}}v_2^3 \left[ \frac{12}{5}X+
 \left(- \frac{24}{7}   X +54   X^3
 -56  X^5 \right)v_2^2\eta^2\right]\,,\nonumber\\
{\mathbf A}^{2.5+3.5}_{y\, \rm PN\, from \, PM}&=& \frac{G^2 m_Am_B}{D_{\rm PN}^3} v_2^3  \left[-\frac45+\frac{12}{5}X^2+\left(\frac{292}{35} - \frac{486}{7}X^2 + 114 X^4- 56 X^6  \right)v_2^2\eta^2\right]\,. 
\eea
\end{widetext}
This result agrees with the result of  Refs. \cite{Iyer:1995rn,Nissanke:2004er,Blanchet:2018yqa} 

Tables \ref{tab:table1x} and \ref{tab:table1y} list the $x$ and $y$ components  of the PN-expansion of our  $O(G^2)$ 
radiation-reaction ${\mathbf A}_A^{\rm rr}$ to a high expansion order. These results might provide useful benchmarks
for future PN-based radiation-reaction computations in harmonic coordinates.

\begin{table*}  
\caption{\label{tab:table1x} ${\mathbf A}^{\rm PN\, from \, PM}_x$ up to 8.5PN, having factored $\frac{G^2m_Am_B}{D_{\rm PN}^3} \frac{b}{D_{\rm PN}}v_2^3$ in front and using the notation \eq{defDX}.}
\begin{ruledtabular}
\begin{tabular}{ll}
2.5PN &$\frac{12}{5}X$\\
3.5PN &$v_2^2\left( - \frac{24}{7}   X+54   X^3-56  X^5 \right)$\\
4.5PN &$ v_2^4 (-\frac{104}{15} X+\frac{578}{7} X^3-\frac{231}{2} X^5+36 X^7) $\\
5.5PN &$ v_2^6 (-\frac{17139}{770} X+\frac{4408}{21} X^3-421 X^5+\frac{1251}{4} X^7-\frac{165}{2} X^9)$\\
6.5PN &$v_2^8 (-\frac{21277}{546} X+\frac{365489}{924} X^3-\frac{6221}{6} X^5+\frac{4707}{4} X^7-\frac{19987}{32} X^9+\frac{2509}{20} X^{11})$\\
7.5PN &$ v_2^{10}(-\frac{131207}{2310} X+\frac{643616}{1001} X^3-\frac{1101433}{528} X^5+\frac{12725}{4} X^7-\frac{40975}{16} X^9+\frac{336193}{320} X^{11}-\frac{1387}{8} X^{13})$\\
8.5PN &$v_2^{12}(-\frac{367429}{4862} X+\frac{11471123}{12012} X^3-\frac{25381357}{6864} X^5+\frac{2494473}{352} X^7-7678 X^9+\frac{306865}{64} X^{11}-\frac{413187}{256} X^{13}+\frac{25415}{112} X^{15})$\\
\end{tabular}
\end{ruledtabular}
\end{table*}

\begin{table*}  
\caption{\label{tab:table1y} ${\mathbf A}^{\rm PN\, from \, PM}_y$ up to 8.5PN, having factored $ \frac{G^2m_A m_B}{D_{\rm PN}^3} v_2^3$ in front and using the notation \eq{defDX} .}
\begin{ruledtabular}
\begin{tabular}{ll}
2.5PN& $-\frac45+\frac{12}{5}X^2$\\
3.5PN& $v_2^2\left(\frac{292}{35}   - \frac{486}{7}X^2+ 114 X^4-56 X^6  \right)$\\
4.5PN& $v_2^4 (\frac{160}{63}-\frac{1844}{105} X^2+\frac{687}{14} X^4-\frac{147}{2} X^6+36 X^8)$\\
5.5PN& $v_2^6 (-\frac{2033}{990}+\frac{89603}{2310} X^2-\frac{337}{6} X^4-\frac{389}{4} X^6+\frac{783}{4} X^8-\frac{165}{2} X^{10})$\\
6.5PN& $v_2^8 (-\frac{14445}{2002}+\frac{2592897}{20020} X^2-\frac{325289}{924} X^4+\frac{503}{3} X^6+\frac{10953}{32} X^8-\frac{65439}{160} X^{10}+\frac{2509}{20} X^{12}) $\\
7.5PN& $v_2^{10}(-\frac{104999}{8190}+\frac{7711687}{30030} X^2-\frac{6355289}{6864} X^4+\frac{1759397}{1584} X^6-\frac{1871}{32} X^8-\frac{293359}{320} X^{10}
+\frac{687869}{960} X^{12}-\frac{1387}{8} X^{14}) $\\
8.5PN& $v_2^{12}(-\frac{316213}{16830}+\frac{431274737}{1021020} X^2-\frac{2721863}{1456} X^4+\frac{135167975}{41184} X^6-\frac{769997}{352} X^8-\frac{112849}{160} X^{10}+\frac{7637851}{3840} X^{12}-\frac{2047417}{1792} X^{14}+\frac{25415}{112} X^{16}) $\\
\end{tabular}
\end{ruledtabular}
\end{table*}

Let us finally mention that we explored the Frenet-Serret geometry of the two worldlines ${\mathcal L}_A$, ${\mathcal L}_B$,
considered as parametrized curves in a Minkowski background. [Such a study is gauge-dependent, and crucially depends on our use
of harmonic coordinates.] We recall that the Frenet-Serret  approach defines the curvature ($\kappa$) and torsions
($\tau_1, \tau_2$) 
of a curve $z^\alpha(s)$, together with an associated moving orthonormal frame $(E_0,E_1,E_2,E_3)$, by successively 
defining  the basis vectors  $(E_0,E_1,E_2,E_3)$ so that
\bea
\label{acc_eq2}
\frac{d z^\alpha}{ds} &=& E_0\,,\nonumber\\
\frac{dE_0^\alpha}{ds} &=& \kappa E_1^{\alpha}\,, \nonumber\\
\frac{dE_1^\alpha}{ds}  &=& \kappa E_0^{\alpha}+\tau_{1} E_2^\alpha\,,\nonumber\\
\frac{dE_2^\alpha}{ds} &=& -\tau_{1} E_1^\alpha+ \tau_{2} E_3^\alpha\,, \nonumber\\
\frac{dE_3^\alpha}{ds} &=& -\tau_{2} E_2^\alpha\,. 
\eea
When applying this approach to  $z^\alpha(s) = z_A^\alpha(\tau_A)$, $E_0$ is $u_A$
and the curvature is just the magnitude of the acceleration  $d u_A^\alpha(\tau_A)/ d\tau_A$. Of more interest
are the geometrical facts that the second torsion $\tau_{2}$ vanishes (because  ${\mathcal L}_A$ lies
within a 3-plane) and that the first torsion  $\tau_{1}$ has a $G$-expansion of the type
\be
\tau_{1}\Big|_{{\mathcal L}_A}=\tau_1^{G^0}+G \tau_1^{G^1}+O(G^2)\,.
\ee
Here, the leading-order $O(G^0)$ term reads
\be
\tau_1^{G^0}= \frac{\Omega}{1+\Omega^2\tau_A^2} \,, 
\ee
with
\be
\Omega=\frac{\sqrt{\gamma^2 - 1} (3-2\gamma^2)\gamma}{(2\gamma^2 - 1) b}\,.
\ee
The value of the  $O(G^1)$ contribution to the torsion is too involved to be displayed here.

We note the interesting fact that the leading-order contribution to the torsion changes sign as the Lorentz factor
$\g$ crosses the value 
\beq \label{g*}
\gamma_*=\sqrt{\frac32}\,.
\eeq
More precisely, $\tau_1^{G^0}$ is positive for $1 <\gamma<\sqrt{\frac32}$,
and negative for  $\gamma>\sqrt{\frac32}$. The critical value Eq. \eq{g*} where the torsion changes
sign is equal to the critical value where the coefficient
\be
 \frac{\gamma (2\gamma^2-3)}{(\gamma^2-1)^{3/2}}
\ee
of the $O(G)$ logarithmic drift of the worldlines (see Eq. \eq{intCG1}) changes sign.

\section{Conclusions}

We have discussed both conservative and radiation-reaction effects on the orbits of two gravitationally interacting massive bodies, at the second Post-Minkowskian order.
In particular,  we have explicitly written the (harmonic-coordinate) solution worldlines, $z_A(\tau_A)$,  $u_A(\tau_A)$, at $O(G^2)$,
both in the conservative and the retarded cases. Several checks of our results have been presented, and notably 
a direct check that the $O(G^2)$ variation of the mechanical angular momentum of the binary system during scattering 
exactly balances the angular momentum radiated at infinity in the form of gravitational waves. 
We also pointed out that the sign of the (flat-background-defined) Frenet-Serret  torsion of the worldlines is
related to the sign of the asymptotic logarithmic drift of the worldlines.

We hope that our explicit worldline results will be useful in future investigations of the gravitational interaction,
and radiation, of binary systems. In particular, let us recall that (in harmonic coordinates)
 the gravitational field ${\frak h}^{\mu\nu} \equiv \sqrt{|g|} g^{\mu\nu}- \eta^{\mu\nu}$ generated by
a binary system  is sourced, 
\be
 \Box {\frak h}^{\mu\nu} = 16 \pi G T^{\mu\nu}_{\rm eff} \, ,
 \ee
 by the effective stress-energy tensor
\be
16 \pi G T^{\mu\nu}_{\rm eff} = 16 \pi G |g| T^{\mu\nu}_{\rm matter}+ \Lambda^{\mu\nu}_{\rm grav}(\frak h)\,,
\ee
where the (Landau-Lifshitz) field contribution $\Lambda^{\mu\nu}_{\rm grav}({\frak h})$ is at least
bilinear in the gravitational field: $\Lambda^{\mu\nu}_{\rm grav}(\frak h) \sim {\frak h} \partial^2 {\frak h} +  ( \partial {\frak h})^2 + \cdots$.
Our  results allows one to write down an explicit {\it $G^3$-accurate} expression for the waveform emitted 
at infinity by the matter contribution
$16 \pi G |g| T^{\mu\nu}_{\rm matter}$ (which is localized on the two worldlines). Inserting in the
field contribution $\Lambda^{\mu\nu}_{\rm grav}({\frak h})$ the $G^2$-accurate explicit expression of 
${\frak h} = G \frak h_1 + G^2 \frak h_2 +O(G^3)$ derived in Ref. \cite{Bel:1981be} then formally yields an explicit {\it $G^3$-accurate}
(spacetime) expression for the field contribution $\Lambda^{\mu\nu}_{\rm grav}(\frak h)$  to $16 \pi G T^{\mu\nu}_{\rm eff}$. 
If one can compute the 
$O(G^3)$ waveform contribution generated by $\Lambda^{\mu\nu}_{\rm grav}(G \frak h_1 + G^2 \frak h_2)$
to complete the waveform emitted by inserting our explicit results in $16 \pi G |g| T^{\mu\nu}_{\rm matter}$, this will
provide a purely classical derivation of the $G^3$-accurate waveform which has been recently derived by quantum methods,
whose application to this problem have raised some subtle issues \cite{Luna:2017dtq,Jakobsen:2021smu,Jakobsen:2021lvp,Mougiakakos:2021ckm,Riva:2022fru,Cristofoli:2021vyo,Adamo:2022qci,Brandhuber:2023hhy,Herderschee:2023fxh,Elkhidir:2023dco,Georgoudis:2023lgf,Gonzo:2023cnv,Aoude:2023dui,Caron-Huot:2023vxl,Georgoudis:2023eke,Georgoudis:2023ozp,Bohnenblust:2023qmy,Georgoudis:2024pdz,DeAngelis:2023lvf,Brandhuber:2023hhl,Bini:2024rsy,Brunello:2024ibk}.

\section*{Acknowledgements}
D.B.  
acknowledges sponsorship of the Italian Gruppo Nazionale per la Fisica Matematica (GNFM)
of the Istituto Nazionale di Alta Matematica (INDAM), 
as well as the hospitality and the highly stimulating environment of the Institut des Hautes Etudes Scientifiques.
The present research was partially supported by the 2021 Balzan Prize for Gravitation: Physical and Astrophysical Aspects, awarded to T. Damour.

\end{document}